%% file: main.tex
\documentclass[sigconf]{acmart}
\acmConference[ICPC 2022]{The 30th International Conference on Program Comprehension}{May 16-17, 2022}{Pittsburgh, PA, USA}
\usepackage{amsmath,amsfonts}
\usepackage{algorithmic}
\usepackage{graphicx}
\usepackage{textcomp}
\usepackage{rotating}
\usepackage{array}
\usepackage{xcolor}
\usepackage{tikz}
\usepackage{makecell}
\usepackage{multirow}
\usepackage{etoolbox}
\def\BibTeX{{\rm B\kern-.05em{\sc i\kern-.025em b}\kern-.08em
    T\kern-.1667em\lower.7ex\hbox{E}\kern-.125emX}}
\begin{document}

\newcommand*\circled[1]{\tikz[baseline=(char.base)]{
		\node[shape=circle,draw,inner sep=0.8pt] (char) {#1};}}
	
\setcopyright{acmcopyright}

\title{Semantic Similarity Metrics for Evaluating \\Source Code Summarization
}





\author{Sakib Haque, Zachary Eberhart, Aakash Bansal, Collin McMillan}
\email{{shaque,zeberhar,abansal1,cmc}@nd.edu}
\affiliation{
	\institution{University of Notre Dame}
	\city{Notre Dame}
	\state{IN}
	\country{USA}
}


\begin{abstract}
Source code summarization involves creating brief descriptions of source code in natural language.  These descriptions are a key component of software documentation such as JavaDocs.  Automatic code summarization is a prized target of software engineering research, due to the high value summaries have to programmers and the simultaneously high cost of writing and maintaining documentation by hand.  Current work is almost all based on machine models trained via big data input.  Large datasets of examples of code and summaries of that code are used to train an e.g. encoder-decoder neural model.  Then the output predictions of the model are evaluated against a set of reference summaries.  The input is code not seen by the model, and the prediction is compared to a reference.  The means by which a prediction is compared to a reference is essentially word overlap, calculated via a metric such as BLEU or ROUGE.  The problem with using word overlap is that not all words in a sentence have the same importance, and many words have synonyms.  The result is that calculated similarity may not match the perceived similarity by human readers.  In this paper, we conduct an experiment to measure the degree to which various word overlap metrics correlate to human-rated similarity of predicted and reference summaries.  We evaluate alternatives based on current work in semantic similarity metrics and propose recommendations for evaluation of source code summarization.
\end{abstract}


\maketitle

\input intro
\input background
\input humanstudy

\input humanresults
\input similarityMetrics
\input corrstudy
\input results

\input discussion

\section*{Acknowledgment}
This work is supported in part by NSF CCF-2100035. Any opinions, findings, and conclusions expressed herein are the authors and do not necessarily reflect those of the sponsors.

\bibliographystyle{IEEEtran}
\bibliography{main}

\end{document}

%% file: intro.tex
\section{Introduction}

Software documentation for programmers is built from source code summaries.  A summary is a brief description of a section of source code that helps programmers understand what the code does, why it exists, etc., without having to resort to reading the code itself.  
Yet, while programmers seek good documentation for themselves, they are notorious for writing poor documentation for others due to time pressure and language barriers~\cite{shi2011empirical, zhong2013detecting}.  A dream of software engineering research for decades is to write this documentation automatically.  Tool support has long focused on formatting and presentation of summaries written in metadata e.g. JavaDocs~\cite{kramer1999api}, but the real prize is to free programmers from the manual effort of writing natural language descriptions at all~\cite{forward2002relevance}.  Research effort towards automatically writing these descriptions has come to be known as source code summarization~\cite{haiduc2010use}.

The current research frontier in source code summarization involves machine learning models trained from big data input (typically this means an attentional encoder-decoder neural model).  The inspiration for these models comes from machine translation.  In machine translation, an ML model is trained using large ``paired datasets'' -- the datasets are ``paired'' in that sentences in one language e.g. French are linked to a close translation of that sentence in another language e.g. English.  The analog in code summarization is that code is paired with summaries of that code.  
These paired datasets serve as training input to models, similar to machine translation.  The hope is that the models will learn to predict novel summaries for code that has not been seen before.

Evaluation of these models is frequently a weak point, as Roy~\emph{et al.}~\cite{roy2021reassessing} argued. 
There are basically two strategies.  One is a data-driven methodology in which a portion of the paired dataset is set aside as a test set.  Once the model is trained, the code in the test set is then shown to it. The model's prediction for that code can then be checked against the reference. 
The trouble lies in how the predictions are ``checked against'' the reference.  By far the most common metric is BLEU~\cite{Papineni:2002:BMA:1073083.1073135}, perhaps combined with ROUGE~\cite{lin2004rouge} or other similar ones.  These metrics measure word overlap between the prediction and the reference.  The advantage is that word overlap is easy to understand and cheap to calculate -- a data-driven evaluation methodology can be easily reproduced and can cover a test set of tens of thousands of examples.  A major disadvantage is that some words in a sentence are more important than others, and many words have close synonyms.  Confusing the word ``add'' for ``delete'' has a big impact on the meaning of a summary.  Confusing ``remove'' for ``delete'' does not.  Yet word overlap metrics treat these mistakes the same.  In addition, many metrics such as BLEU are intended as corpus-level metrics, and tend to be unreliable indicators at finer granularity such as sentence-level~\cite{reiter2018structured}.

In software engineering research, the other evaluation strategy is a human study.  Feedback from human experts is widely viewed as the gold standard in any evaluation.  The advantages seem very clear: expert human judgment is correct almost by definition in these studies.
Humans can provide deep insights into the correctness of model predictions as well as their usefulness for a particular task, includes a specific type of information, etc.  
But studies with humans also have disadvantages.  The vicissitudes of human life include fatigue, technical errors, and a multitude of biases.  Human time is also expensive.  A person may take several hours to evaluate just a few hundred examples.  The result is that the view of prediction quality has depth but often lacks breadth, and these studies are often impossible to replicate. Therefore, few experiments combine automated metrics like BLEU with human evaluation. As noted by Stapleton \emph{et al.}~\cite{stapleton2020human}, human studies are very rare in practice.

One alternative proposed in other domains is measurement of semantic similarity.  
Semantic similarity metrics work by modifying word overlap metrics to consider how similar words are in an embedding space of those words.  Weiting~\emph{et al.}~\cite{wieting2019beyond} call this ``partial credit.''  The idea is that instead of assigning a one or zero for a word hit or missed, the metric should assign a score based on the semantic similarity of that word.  This semantic similarity can be calculated in any number of ways, such as vector similarity in a pretrained word embedding.  Semantic similarity for measuring code summaries is not a novel concept 
as evidence is accumulating that word overlap measures are not sufficient~\cite{9286030, stapleton2020human}.  Yet the literature does not clearly recommend which semantic similarity metric should be used as an alternative, and how to use it for code summarization.

In this paper, we study word overlap and semantic similarity metrics for evaluating source code summarization.  We perform an experiment in which we asked human experts to evaluate the quality of several hundred predictions from a baseline neural source code summarization technique.  We also asked (different) human experts to evaluate the quality of the reference summaries themselves, and their perceived similarity of the predicted summaries to the references.  We then compute correlation of this perceived similarity to several word overlap and semantic similarity metrics from the related NLP literature.  Using our findings from this experiment, we provide recommendations about which similarity metric best correlates to similarity as perceived by the human experts.  We provide recommendations for using this and other metrics in research on code summarization. 

Results from our experiments showed that the Sentence Bert Encoder (SentenceBert)~\cite{reimers2019sbert} produced vectorized representations of the summaries that have the highest correlation to perceived similarity by the human experts in our experiment.  The Spearman correlation of SentenceBert was approximately 14\% higher than BLEU.  In fact, BLEU ranked near the bottom in terms of correlation to human-perceived similarity of source code summaries, when compared to several  
sentence embeddings.  These newer embeddings also scored higher in terms of correlation to human-perceived accuracy, completeness, and conciseness, which are widely-recognized criteria for documentation quality.  This result is significant because BLEU is by far the most common metric used when evaluating source code summarization.  Our result implies that researchers in the area of code summarization should use SentenceBert or another similar metric in addition to BLEU. We do not view our work as a “be all end all” solution. Rather, we view our work as one piece of evidence contributing to a growing body of literature surrounding evaluation methodologies for source code summarization.

To maximize reproducibility and usefulness to other researchers, we release all source code and data (see Section~\ref{sec:reproducibility}).  

%% file: background.tex
\section{Background \& Related Work}
\label{sec:background}

In this section, we discuss the current status of source code summarization research, as well as similarity metrics.

\subsection{Source Code Summarization}

This history of source code summarization can be broadly classified into two groups: 1) heuristic/template-based, and 2) data-driven.  Heuristic and template based approaches include work by Sridhara~\emph{et al.}~\cite{sridhara2011automatically, sridhara2011generating}, McBurney~\emph{et al.}~\cite{mcburney2016automatic}, and Moreno~\emph{et al.}~\cite{moreno2014automatic}.  The basic idea behind these approaches is to use manually-encoded rules to detect patterns in the source code associated with particular types of comments, and then extract information from those patterns for use in predefined templates.  Around the same time, heuristics based on Information Retrieval (IR) were also popular~\cite{haiduc2010use, de2012using, rastkar2014automatic, rodeghero2014improving}.  The common theme to these approaches was to use IR to extract a set of words or sentences from software artifacts that explains those artifacts.  For example, term frequency / inverse document frequency was a popular metric for choosing a top-$n$ word list for software components such as Java methods.

Around 2016, the winds changed strongly in the direction of neural networks, and in particular the attentional encoder-decoder neural architecture.  These approaches are data-driven in that they rely on big data input, such as large source code repositories~\cite{leclair2019recommendations}.  Approaches in this vein are now too numerous to list exhaustively~\cite{iyer2016summarizing, jiang2017automatically, hu2018deep, alon2019code2seq, alon2019code2vec, leclair2019neural, leclair2020improved, ahmad2020transformer, haque2020improved}.  The common element to these approaches is that they take source code as input in various formats (text token, AST representation, etc.) and learn to create a vectorized representation of source code via one technique or another (RNN, GNN, Transformer, etc.). They then use this vectorized representation to predict words for a source code summary.

The shift to data-driven approaches also brought a shift to data-driven evaluation. Roy~\emph{et al.}~\cite{roy2021reassessing} present a critique of this shift by demonstrating that small improvements in metrics are not necessarily borne out in human evaluations.  Heuristic and template-based approaches almost always needed evaluation by human expert opinion because there was no consistent gold set against which to evaluate -- the template, for example, could result in a high quality summary even if it did not match the reference summary.  In contrast, data-driven approaches are by definition trained with large datasets.  These datasets are usually split into training/test/validation subsets, in which the validation and test sets may each be 5-10\% of the dataset.  The result is a test set with tens of thousands of examples.  This large size is well beyond what a human can be expected to evaluate by hand.  The solution came in the form of word overlap metrics borrowed from the NLP domain.

\subsection{Measuring Word Overlap}

By far the most popular metric for measuring word overlap in source code summarization research is BLEU~\cite{Papineni:2002:BMA:1073083.1073135}.  BLEU originated in 2002 from the machine translation research community where it was intended to measure predicted translations' similarity to reference translations.  E.g. a predicted English translation of a French sentence would be compared to a reference English translation.

BLEU works by comparing n-grams in the predicted and reference summaries.  In the most typical implementation, $n$ ranges from 1 to 4 and is used to compute a BLEU$_n$ score:

\[ BLEU_n = \frac{\sum_{t_n} \min \{ C_p(t_n), C_{r}(t_n) \}}{P(n)} \]

Where $t_n$ is an $n$-gram in the prediction, $C_p(t_n)$ is the count of that $n$-gram in the prediction, $C_r(t_n)$ is the count of that $n$-gram in the reference, and $P_n$ is the total number of $n$-grams (for that size $n$ only) in the prediction.  Note that the metric BLEU$_1$ is identical to unigram precision; it is the overlap of words regardless of the order of those words.

Usually a single aggregate ``BLEU score'' is reported, which is basically the product of the BLEU$_n$ scores. 
Sometimes slight modifications are used, such as adding more weight for BLEU$_n$ scores with higher values of $n$ because a single correct 4-gram is considered better than four correct unigrams.  
Also, a brevity penalty is often applied to penalize the model for generating very short predictions.  Additionally, the original definition of BLEU allows for multiple correct references to be provided, though in code summarization research there is almost always only one reference.  Due to the high number of variants and parameters, we use the BLEU package implemented by the popular NLTK {\small \texttt{nltk.translate.bleu\_score}}~\cite{NLTKWebsite}.

Another issue is that BLEU is intended as a \emph{corpus-level} metric, not sentence-level.  BLEU is intended to provide a big picture view of a model's performance over the entire test set, and may not be a reliable indicator of performance on any given data entry.  Indeed, Reiter~\cite{reiter2018structured}, in a meta-review of papers with both BLEU and human studies, found that only corpus-level BLEU correlated with the results in the human studies.  Yet in practice, what programmers read are individual summaries -- it is the performance on a specific summary that matters to a human reader.  This lack of reliability at the sentence-level is a long-standing complaint against BLEU among researchers in several fields~\cite{novikova-etal-2017-need, van2019best, clark2019sentence}.

Alternatives (notably ROUGE~\cite{lin2004rouge} and METEOR~\cite{banerjee2005meteor}) have been proposed to address specific complaints about BLEU, and we provide more details about the metrics we use in this paper in Section~\ref{sec:simmetric}.  But the point is that while measures of word overlap are easy to calculate, they all basically operate by measuring whether the same words are used in the same order in both predictions and references.

\subsection{Measuring Semantic Similarity}

Measures of semantic similarity have long been proposed as improvements to word overlap.  Semantic similarity is defined as a similarity between two texts in terms of what those texts actually mean.  We leave details of these approaches to Section~\ref{sec:simmetric}, where we discuss each of the metrics we use.

The basic idea behind most current semantic similarity metrics is to use a sentence encoder to produce a vectorized representation of two sentences, and then measure the cosine or Euclidean similarity of the vectors representing those sentences.  The sentence encoding techniques vary in terms of their strategy and underlying training data, but the idea is to learn an encoding from a large dataset e.g. via as a language model.  These encodings can apply ``partial credit''~\cite{wieting2019beyond} for synonyms, misspellings, etc., based on how words are used in the underlying training data.  The vectorized representation of a sentence will hopefully be similar to that of another sentence if they have words that tend to be used the same way, rather than relying only on including the same words in the same order.


%% file: humanstudy.tex
\newpage
\section{Empirical Study with Programmers}
\label{sec:humexp}

We perform an empirical study with human programmers as a prelude to studying semantic similarity metrics.  This human study provides data against which semantic similarity metrics can be evaluated, as well as human subjective ratings of sample source code summaries.

\subsection{Research Questions}

The research objective of this study is to determine the level of semantic similarity between source code summaries, and the level of perceived overall quality of those summaries.  To that end, we ask the following Research Questions (RQs):

\begin{description}
	\item[RQ$_1$] What is the perceived level of quality of summaries from a code summarization baseline approach?
	\item[RQ$_2$] What is the perceived level of quality of the reference source code summaries?
	\item[RQ$_3$] What is the perceived level of semantic similarity between the generated and reference summaries?
\end{description}

The purpose of RQ$_1$ is to determine a baseline level of quality for source code summaries predicted by a typical neural code summarization approach.  At present, nearly all evaluations of code summarization approaches have been performed using BLEU or other word overlap metrics.  These calculate similarity to a reference, but not overall quality of the summaries themselves.  In contrast, humans are able to provide a subjective rating of quality along different axes (e.g. conciseness, completeness, accuracy) that is independent of the overlap to the reference.  We aim to collect this baseline level of quality for comparison to semantic similarity measures, but also for general academic interest in understanding how well a current popular baseline approach is perceived to work.

The rationale behind RQ$_2$ is that data-driven code summarization approaches are trained using reference examples from source code repositories, but the quality of these references is largely unknown.  Different strategies have been proposed to attempt to ensure adherence to accepted good practice~\cite{leclair2019recommendations, allamanis2019adverse}, but the literature does not describe studies of the quality itself.  This is a problem because the reference examples form a ceiling above which most existing code summarization approaches will not perform -- after all, these approaches are trained on the reference examples.

The purpose of RQ$_3$ is to determine how similar human programmers perceive generated source code summaries to be to reference summaries.  We use these perceived similarity ratings as a gold set against which we evaluate calculated similarity metrics in the Section~\ref{sec:semexp}.

\subsection{Methodology}

Our methodology is inspired by the accepted practice for evaluating source code summarization approaches described by McBurney~\emph{et al.}~\cite{mcburney2016automatic} and Sridhara~\emph{et al.}~\cite{sridhara2010towards}.  In a nutshell, this practice involves asking questions related to three quality criteria: overall accuracy, completeness, and conciseness.  The idea is that a summary should have only correct information, should not be overly verbose, and yet should include enough information to understand what the code does.  

\begin{figure}[b]
	\vspace{-0.1cm}
	\centering
	\includegraphics[width=0.49\textwidth]{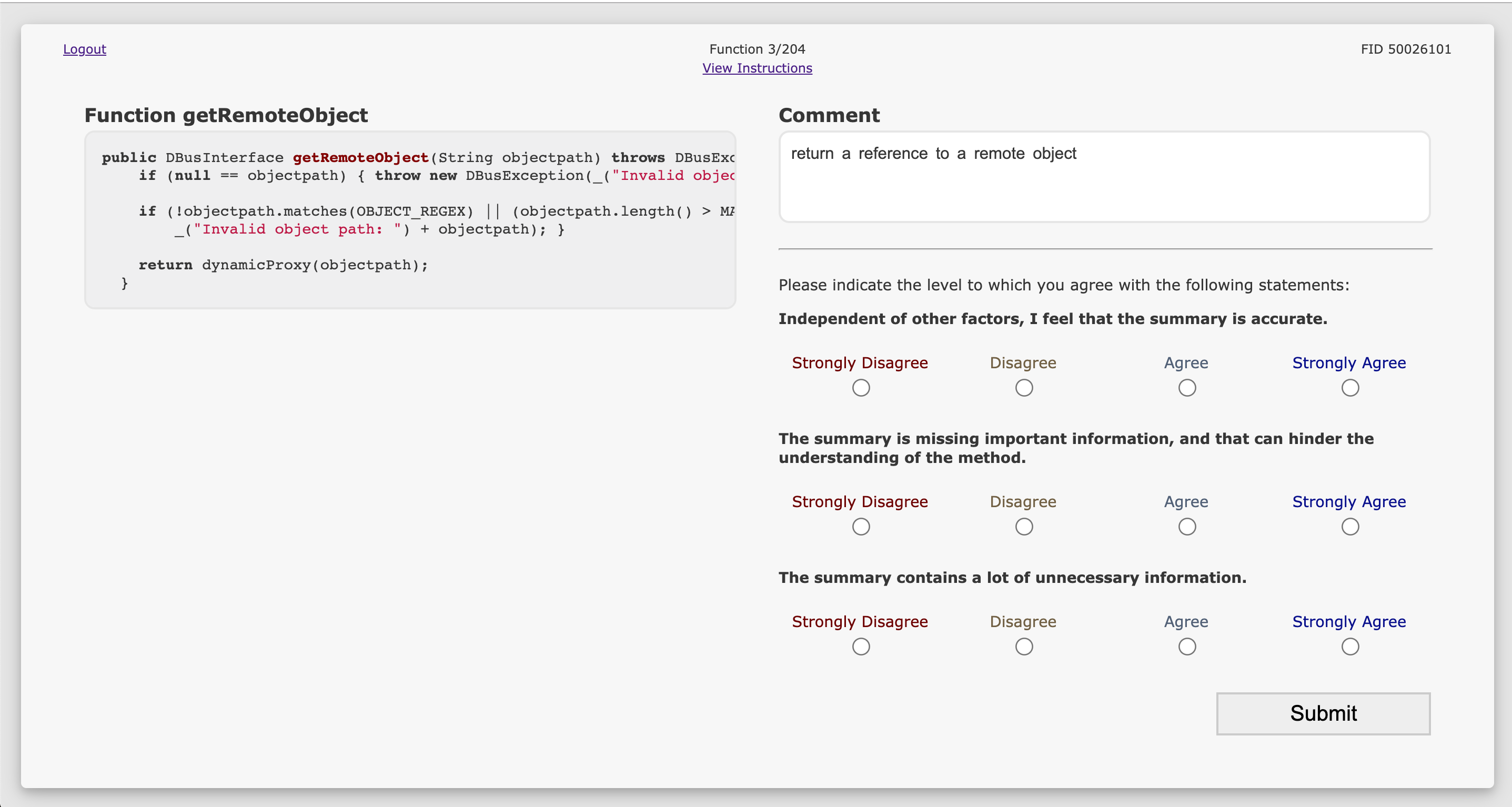}
	\vspace{-0.5cm}
	\caption{A screenshot of the interface used by participants during the human study.  The source code of a subroutine is display alongside a summary of that subroutine.  Questions are shown to the right.}
	\label{fig:webinterface}
\end{figure}

We recruited participants for a web survey in which they viewed source code summaries and answered four questions about those summaries.  Participants had no contact with one another and did not know how many others were participating.  When a participant activated the web survey, the survey displayed the interface shown in Figure~\ref{fig:webinterface}.  The source code of a subroutine is displayed along with a summary of that subroutine.  Then on the left, the participant evaluates the summary by answering three questions:

\begin{table}[!h]
	\vspace{-0.2cm}
	\begin{tabular}{lp{7.5cm}}
		Q$_1$ & Independent of other factors, I feel that the summary is accurate.                                 \\
		Q$_2$ & The summary is missing important information, and that can hinder the understanding of the method. \\
		Q$_3$ & The summary contains a lot of unnecessary information.                                            
	\end{tabular}
	\vspace{-0.3cm}
\end{table}

The participant could choose one of four options for each question, ranging from ``Strong Disagree'', ``Disagree'', ``Agree'', and ``Strongly Agree.''  The procedure so far is as recommended by McBurney~\emph{et al.}~\cite{mcburney2016automatic}. The questions are necessary (as opposed to simply asking for ``accuracy'') to help ensure participants have a consistent understanding of the prompt, as Novikova~\emph{et al.}~\cite{novikova-etal-2017-need} also note in a study of BLEU.  A difference is that we also need to collect perceived similarity of generated and reference summaries.  So, after the participant answers the three questions above and clicks ``next'', the interface shows the different summary in addition to the one already shown, and ask a fourth question:

\begin{table}[!h]
	\vspace{-0.1cm}
	\begin{tabular}{lp{7.5cm}}
		Q$_4$ & These two comments are similar.                               \\
    \end{tabular}
	\vspace{-0.5cm}
\end{table}

The participant selects one of the same four options above.

The interface continues this process for 210 subroutines.  We chose 210 because this is what we found that most participants could reasonably be expected to complete in around 3.5 hours (a rate of approximately one minute per subroutine).  The interface shows the 210 subroutines in random order, but all participants completed all 210 evaluations.  The interface chose at random whether to show a participant a generated or a reference summary for evaluation under Q$_1$-Q$_3$.  If the generated summary was shown, then the reference would be shown for Q$_4$.  If the reference was shown, the generated would be shown for Q$_4$.  But, the interface ensured a balance between showing the generated or reference summary for Q$_1$-Q$_3$, so each summary would be evaluated by half the participants.

\subsection{Subject Participants}

We recruited thirty professional programmers to participate in this study.  These programmers were all Java developers by trade in a variety of industries (redacted due to privacy policy). On average, the participants had $9.3$ years of Java programming experience with a median experience of $8.5$ years. We recruited these programmers via Upwork, offering remuneration of US\$60/hr, market rate in our location. We did not use Amazon Mechanical Turk (AMT)~\cite{paolacci2010running} because Eberhart \emph{et al.}~\cite{eberhart2019automatically} shows that AMT users demonstrate lower similarity in agreement. They argue that this is due to lack of expert domain knowledge and we need high degree of reliability to recommend the use of a new metric.

\subsection{Baseline Code Summarization}

We used the baseline neural source code summarization technique called {\small \texttt{attendgru}} to create the generated source code summaries in this study.  This baseline is essentially a vanilla attentional encoder-decoder seq2seq-like neural model, using a single GRU for the encoder and another GRU for the decoder.  While it is no longer the state of the art, we use this baseline for two reasons: First, it is well-understood and has been used extensively as a strong baseline in numerous papers~\cite{leclair2019neural, ahmad2020transformer, alon2019code2seq}.  Second, a vast majority of neural code summarization approaches are based on this, or a similar variant of this model~\cite{haque2020improved,hu2018deep}.

\subsection{Subject Code Summaries}
\label{dataset}
We chose 210 Java methods from the test set in the data presented by LeClair~\emph{et al.}~\cite{leclair2019recommendations} as the subject code summaries in this study.  We chose that dataset because it is vetted in peer-reviewed literature and is gaining some traction as a standard set for code summarization experiments.  The test set includes over 90k Java methods, from which we randomly select 210 for this human study.

\subsection{Threats to Validity}

Key threats to validity include: 1) selection of code summaries, 2) selection of baseline, 3) selection of subject participants, and 4) our interface design.  We selected Java methods at random from the test set of a large dataset for code summarization.  While the random selection is hoped to capture a representative sample, it is possible to accidentally create a ``lucky'' or ``unlucky'' random selection.  In this case, the results from the experiment may vary if the random selection happens to include some functions for which the reference or generated summaries are better or worse than average.  The selection of the baseline code summarization approach is also a threat to validity.  It is possible that a different baseline would lead to different conclusions in this study.  We attempted to mitigate this risk by using a popular baseline with ``no frills'' that could increase the number of experimental variables, but the risk still remains that a different approach would lead to different conclusions.  The study participants, like any in any human study, are a threat to validity because different participants may give different answers.  We attempted to mitigate this threat by choosing participants from a worldwide pool and from many industries, to diversify the set of experiences.  A final risk is our interface design.  We used a plain interface in line with previous, similar experiments, but the risk still remains that a different interface could lead to different results.

%% file: humanresults.tex
\section{Programmer Study Results}
\label{sec:progresults}

We answer research questions RQ$_1$-RQ$_3$ in this section.  We also discuss factors related to reliability of our results.

\subsection{RQ$_1$: Perceptions of Generated Summaries}

\begin{figure}[!b]
	\centering
	\includegraphics[width=8cm]{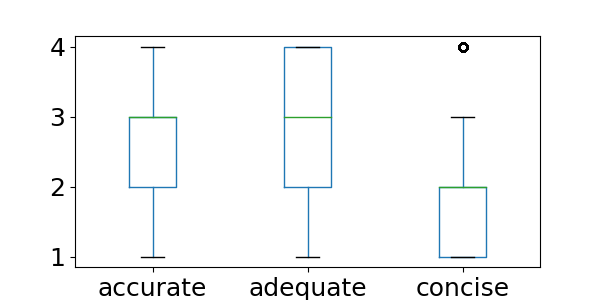} \\
	{\small (a) RQ$_1$: Machine generated.} \\
	\vspace{0.2cm}
	\includegraphics[width=8cm]{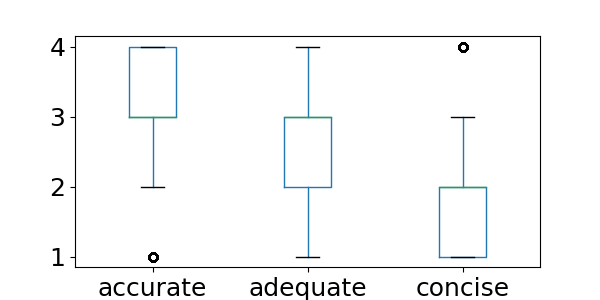} \\
	{\small (b) RQ$_2$: Reference.}
	\vspace{-0.1cm}
	\caption{Boxplots of human ratings for accuracy, completeness, and conciseness.  Plots for RQ$_1$ (a) include ratings for machine generated summaries.  Plots for RQ$_2$ (b) include ratings for reference summaries.  1 = Strongly Agree, 2~=~Agree, 3 = Disagree, 4 = Strongly Disagree (lower is better).}
	\label{fig:rq12boxplots}
	\vspace{-0.1cm}
\end{figure}

We found that the participants in our user study perceived the summaries generated by the baseline code summarization approach to tend to lack accuracy and completeness, while being relatively concise.  Figure~\ref{fig:rq12boxplots}(a) depicts the ratings as boxplots.  Recall that a rating of 1 is ``Strongly Agree'' to 4 is ``Strongly Disagree'', to statements in support of each quality criterion.  A mean score of 1 for e.g. accuracy would mean that the participants strongly agree that the summaries are accurate.  What we observe is that the mean level of accuracy is about 2.64, while the mean level for completeness is about 2.85.  The message is that the study participants tended to disagree that the summaries were accurate and complete.  Meanwhile, the summaries tended to be perceived as concise, with a mean of around 1.88.

\begin{figure}[!b]
	\vspace{-0.4cm}
	\centering
{\scriptsize
\begin{verbatim}
String buildSpatialQueryString(String fullTextQuery,
                               Float latitude,
                               Float longitude,
                               Float radius)
{
    String queryString = "{!spatial circles=" + 
                          latitude.toString() + "," +
                          longitude.toString() + "," +
                          radius.toString() + "}" +
                          fullTextQuery;
    return queryString;
}
\end{verbatim}
\vspace{-0.2cm}
{\small (a) Source code of example Java method.}
\vspace{0.4cm}
}

	{\small
	\begin{tabular}{l|p{5cm}}
		generated     & build a query from the given string     \\
		reference     & build the query string for a spatial query using spatial solr plugin syntax     \\
	\end{tabular}
}

\vspace{0.3cm}
{\small (b) Summaries for example method.}
\vspace{0.4cm}

{\small
	\begin{tabular}{l|ll}
		& generated & reference \\ \hline
		accuracy     & 1.5       & 2.5       \\
		completeness & 4         & 3       \\
		conciseness  & 2         & 3        \\
	\end{tabular}
}

\vspace{0.3cm}
{\small (c) Mean user ratings for example method.}

\caption{Example Java method, summaries, and user ratings.  Note higher perceived accuracy, but lower completeness scores for generated versus reference summary.}
\label{fig:example1}

\end{figure}

\subsection{RQ$_2$: Perceptions of Reference Summaries}

We found that the participants also perceived the reference summaries to lack accuracy, though they tended to be more complete.  The mean accuracy for reference summaries was around 3.05, while completeness was around 2.47.  We note in Figure~\ref{fig:rq12stattest} that the differences in accuracy and completeness between generated and reference summaries are statistically significant as calculated by a Mann-Whitney U test.  In other words, study participants tended to view reference summaries as having more useful information, but that the information that generated summaries did include tended to be more accurate.  To understand this result, consider the example in Figure~\ref{fig:example1}.  The baseline code summarization approach has a tendency to write summaries that rephrase the words in the method signature e.g. ``build a query'' from the method name {\small \texttt{buildSpatialQueryString}}.  This tends to result in accurate summaries that may lack insights provided in the references, such as the ``solr plugin syntax.''  This finding verifies related work~\cite{haque2021action}.  Our point here is that an aggregate score alone (e.g. BLEU) can miss these nuances of human perception.



\begin{figure}[!b]
	\vspace{-0.2cm}
	\centering
	\small
	\begin{tabular}{l|l|l}
					 & U         & p                \\ \hline
		accuracy     & 3826858  & \textless{}0.0001 \\
		completeness & 3934850 & \textless{}0.0001 \\
		conciseness  & 4688546  & \textless{}0.0001           
	\end{tabular}
	\caption{Mann-Whitney U test components and results comparing ratings for generated and reference summaries for each quality criteria.}
	\label{fig:rq12stattest}
	\vspace{-0.3cm}
\end{figure}

\begin{figure}[!b]
	\centering
	{\small
		\begin{tabular}{lll}
			\multirow{4}{*}{\includegraphics[width=2cm]{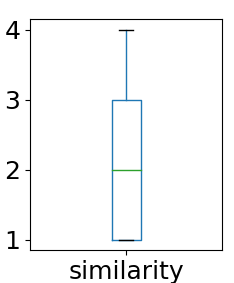}} & mean     			& 2.160 \\
			& geometric mean 	& 1.914 \\
			& harmonic mean 	& 1.688 \\
			& median  			& 2     \\
			& stddev   			& 1.021 \\
			& cv       			& 0.473 \\
			& variance 			& 1.043 \\
			& ~ & ~ \\
			& ~ & ~ \\
		\end{tabular}
	}
	\vspace{-0.7cm}
	\caption{Summary of human-reported similarity between generated and reference source code summaries for RQ$_3$.}
	\label{fig:rq3data}
\end{figure}

\subsection{RQ$_3$: Perceived Similarity of Summaries}

We found that users tended to agree that generated and references are similar (arithmetic mean around 2.2, median 2), though this agreement is not that strong. More than half of the ratings of similarity were either 1 or 2, ``Strongly Agree'' or ``Agree.''  Of the 3 and 4 ratings, more tended to be 3 than 4.  We do not draw strong conclusions from this result.  We report this information merely to show the range of perceived similarity, which we use in the next section to compute correlation to semantic similarity.

\subsection{Aggregating Human Ratings}

In this study, we create a single ``aggregate human-rated score'' for each metric on each summary.  This score is the mean of each of the human rated scores.  Recall that we asked the participants to rate the accuracy, completeness, conciseness, and similarity of code summaries on a scale from 1 to 4.  For each code summary, the aggregate human-rated score for e.g. accuracy is the mean of all accuracy scores given for that summary.  This aggregate score is necessary because different people are likely to have slightly different opinions, which may lead to noise in the data even if the people are in general agreement with each other.

Our choice to aggregate human ratings is in line with procedure established by McBurney~\emph{et al.}~\cite{mcburney2016automatic} and Wood~\emph{et al.}~\cite{wood2018detecting} by reporting facts about agreement, even if taking no action to force agreement by the professional programmers participating in our study.  In general, we found that paired participants gave exactly the same rating approximately 45\% of the time.  Yet it does not mean that the other 55\% are invalid. Consider:



\begin{figure}[!h]
	\vspace{-0.4cm}
	\centering
	\includegraphics[width=6cm]{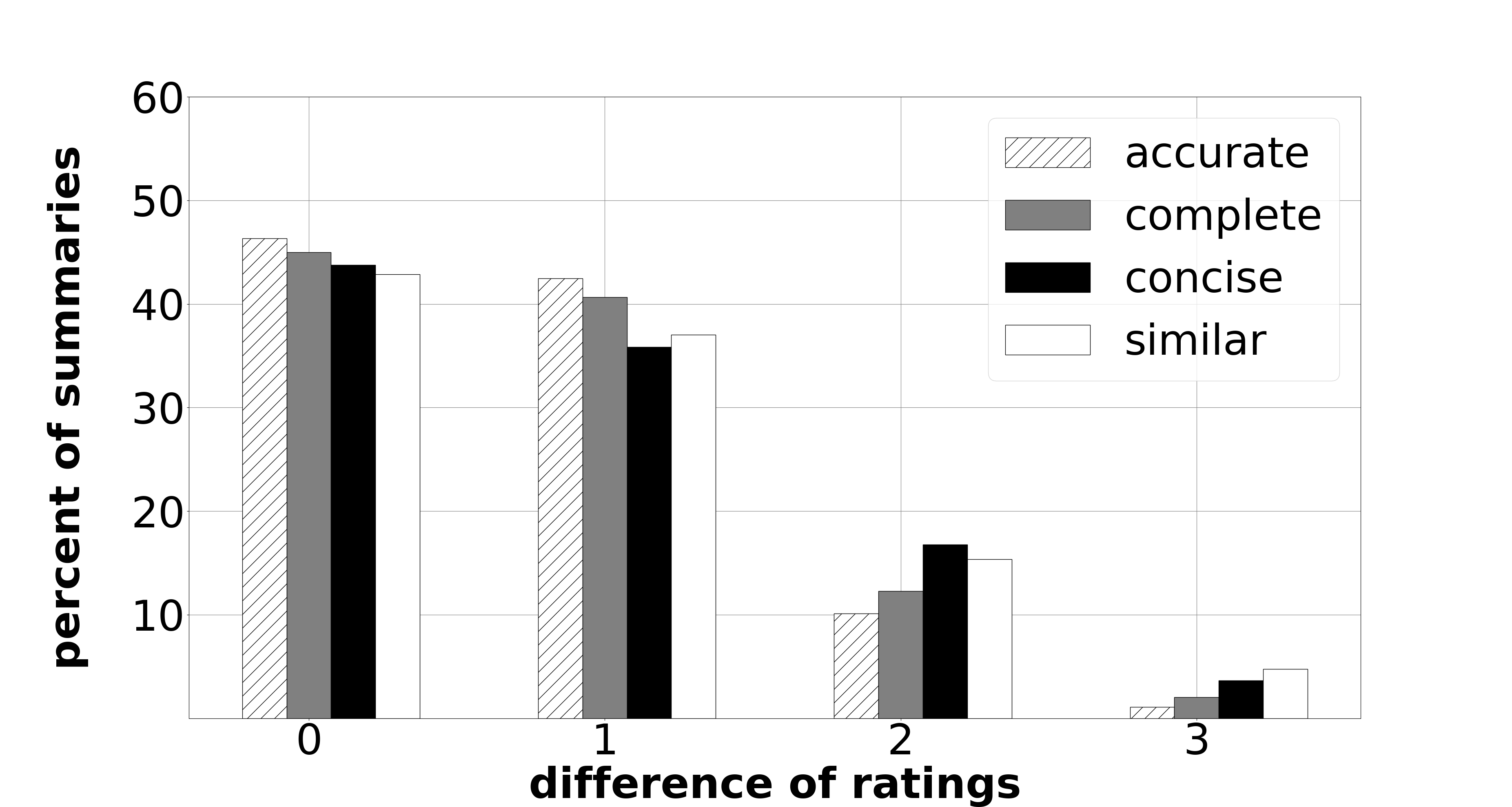}
	\vspace{-0.3cm}
	\vspace{-0.1cm}
\end{figure}

This figure shows the difference between ratings for the same summaries between each participant pair.  For example, if we randomly selected two participants rating the summary similarities, they gave the same rating (zero difference) 46\% of the time and differed by one (e.g. one user selected agree while another said strongly agree) 42\% of the time. Only in 11\% instances were the ratings different by two or three.

Any study with human participants will raise a question about the reliability of the ratings provided by those participants.  Usually the term ``reliability'' refers to how consistent the ratings given by different people are~\cite{krippendorff2011agreement, hsieh2005three}.  Sometimes measuring reliability is straightforward, such as calculating a metric such as Krippendorff's alpha or Cohen's kappa~\cite{kraemer2014kappa}.  Yet these metrics are very controversial, and even their designers offer many caveats~\cite{krippendorff2004reliability}.  A growing consensus is that individual rater's opinions often all have value, that disagreements often should not merely be decided by majority vote, and that studies where raters disagree often reflect valid diversity of opinion and not lack of reliability~\cite{craggs2005evaluating, sun2011meta, potter1999rethinking}.

Therefore, in this paper, we create the aggregate human-rated score as a compromise between forcing participants to agree (such as by engaging in conversation with each other until agreement is reached~\cite{eberhart2020wizard}) and removing results that disagree.  
We use this aggregated human-rated score for research questions RQ$_4$ through RQ$_6$, where we compare semantic similarity metrics.  We did not use this score for RQ$_1$ through RQ$_3$, where we only report human ratings without comparing them to semantic similarity metrics.

%% file: similarityMetrics.tex
\section{Similarity Metrics}
\label{sec:simmetric}

In this section, we describe the similarity metrics we use in this paper.
We broadly categorize these metrics under two approach types, using the same categorization as Zhang et. al.~\cite{zhang2019bertscore}: 1) n-gram matching , and 2) similarity between word or sentence embedding.

Note that in this paper, we are primarily interested in \emph{sentence-level} similarity.  We calculate all metrics at sentence-level, even if they are usually intended as corpus-level metrics (chiefly, BLEU).  Results in this paper should be considered valid only for sentence-level evaluation.

\subsection{N-gram matching}
An n-gram matching metric compares the count of n-grams in the prediction output against the reference output.
High n-gram overlap indicates that the predicted and reference outputs have more common words in order. 

We study the following n-gram matching metrics:

\textbf{Jaccard similarity} is a measure of unigram overlap~\cite{jaccard1912distribution} ~\cite{tanimoto1958elementary}.
Jaccard similarity models each sentence as a set of the words in that sentence.
It is the size of the intersection of those sets, divided by the size of the union of those sets.
An advantage of Jaccard is that it is an intuitive measure of finding common words and assessing senetence similarity.
However, it cannot capture word order.
The sentence pairs ``dog bites man" and ``man bites dog" will generate a Jaccard coefficient of $1$, even though the two sentences are semantically opposite.

\textbf{BLEU} is a precision based measure of n-gram overlap~\cite{Papineni:2002:BMA:1073083.1073135}.
It compares n-grams in the prediction and reference outputs.
An average BLEU score is computed by combining BLEU$_{1}$ to BLEU$_{n}$ using predetermined weights for each $n$.
A more detailed discussion on BLEU can be found in section~\ref{sec:background}.
For our evaluation, we report BLEU$_{1}$ and the average BLEU score with $n$ ranging from 1 to 4 and weights $0.25$ for each $n$.

\textbf{ROUGE} is a recall oriented measure of n-gram overlap between prediction and reference output~\cite{lin2004rouge}.
ROUGE has 4 components: ROUGE-N($R_n$), ROUGE-L($R_l$), ROUGE-W($R_w$) and ROUGE-S($R_s$).
We report $R_l$ and $R_w$ in this paper.
$R_l$ is an f-measure based on the longest common subsequence (LCS) between prediction and reference output.
It is calculated using the following equation:
\[ R_l = \frac{(1+\beta^2)C_{l}P_{l}}{C_{l}+\beta^2P_{l}} \]

Where $C_{l}$ is the LCS based recall and $P_{l}$ is the LCS based precision. 
These are computed as follows:
\[ C_l = \frac{len(LCS(O_p, O_r))}{len(O_r)}, P_l = \frac{len(LCS(O_p, O_r))}{len(O_p)} \]

For reference output $O_r$ and prediction output $O_p$.
$R_w$ introduces a penalty for n-grams in the LCS that are not consecutive.

\textbf{METEOR} combines unigram-precision and unigram-recall to compute a measure for sentence similarity~\cite{banerjee2005meteor}.
Similar to BLEU and ROUGE, METEOR tries to make an exact match between unigrams.
However, if an exact match cannot be made, it tries to match word stems for the remaining words instead.
If a match still cannot be found, it looks to match word synonyms.
Next, it computes the unigram precision and recall followed by the harmonic mean between them.
Finally, it applies a penalty that decreases with higher n-gram matching to compute the final METEOR score.
\input embeddingMetrics

%% file: embeddingMetrics.tex
\vspace{-0.4cm}
\subsection{Similarity of Word or Sentence Embedding}
There are basically two types of semantic similarity metric: 1) word-based similarity, and 2) sentence-based similarity.  
Word-based similarity includes approaches like LSS~\cite{croft2013fast} and STATIS~\cite{li2006sentence} based on labeled relationships among words such as WordNet~\cite{fellbaum2010wordnet}, as well as approaches based on word embeddings learned from large repositories of text~\cite{kenter2015short}.
A word embedding is a fixed-length representation of words in a vector space~\cite{mikolov2013efficient}~\cite{pennington2014glove}.
Word embeddings were developed to find numerical representation of words from text where words with similar meaning have similar representation.
This made neural based learning for tasks using natural language data more efficient~\cite{mikolov2013efficient}.

Sentence embeddings have since been developed to learn vector representation of continuous texts~\cite{le2014distributed}.
These are fixed-length, real-valued representations of entire sentences and their semantics as dense vectors.
Sentence embedding vectors help to understand the context of the words in a sentence.
Pre-trained sentence embeddings also show higher performance in transfer learning tasks than using pre-trained word embeddings~\cite{conneau2018supervised}.
Sentence-based similarity metrics, such as BertScore~\cite{zhang2019bertscore}, use these embeddings to compute the closeness between sentences.

Consider the following sentence pairs taken from the dataset discussed in section~\ref{dataset}: ``gracefully shutdown the application" and ``exits the event."
These sentences have only one unigram matching (``the") and no higher n-gram matching.
Unsurprisingly, these sentences have poor scores across all the n-gram matching based techniques discussed earlier.
However, while they may not have a lot of common words, they are clearly very similar to each other in meaning. 
Therefore, any sentence embedding should have the two sentences close to each other in vector space.

In this paper, we report two different measures of similarity for various embeddings: cosine similarity and Euclidean distance.
Cosine similarity is an angular measure while Euclidean distance is a spatial measure of how close two vectors are in an n-dimensional vector space.
Cosine similarity is calculated by taking the inner product of the two normalized vectors.
Euclidean distance is found by calculating the l2-norm of the two vectors.
Notice that a high cosine similarity score indicates a higher similarity while a higher euclidean distance indicates a lower similarity between sentences.

We explore the following embedding space in this paper:

\textbf{BertScore}~\cite{zhang2019bertscore} measures sentence similarity using BERT based embeddings~\cite{devlin2019bert}~\cite{liu2019roberta}.
BertScore represents each sentence as a sequence of tokens where each token is a word in the sentence.
It then uses a pre-trained contextual embedding of different variants of BERT (we use a 24-layer RoBERTa$_{large}$ model) to represent the tokens.
Next it computes the pairwise inner product (pre-normalized cosine similarity) between every token in the reference and predicted output.
Finally it matches every token in the reference and predicted output to compute the precision, recall and F1 measure.
In this paper, we report the F1 measure of BertScore.

\textbf{Term frequency-inverse document frequency (tf-idf)} reflects the importance of a word to the entire document~\cite{salton1988termweighting} \cite{ramos2003usingtfidf}.
$TF$ reflects how often a word appears in a document. 
$IDF$ weighs up the effect of infrequent words and adjusts for commonly occurring words across all documents like `the', `to', `are', etc. 
The final tf-idf score for each word is calculated by multiplying the values for TF and IDF.
We represent each sentence as a vector of tf-idf score for every word and compute the similarity metrics of corresponding reference and predicted outputs.

\textbf{InferSent} uses a siamese neural network architecture to produce sentence embeddings~\cite{conneau2018supervised}.
It takes a pair of sentences and uses GloVe vectors as pre-trained word embeddings for the sentence pair~\cite{pennington2014glove}.
These embeddings pass through identical RNN encoder layers to produce a fixed-length vector representation of each sentence.
For the encoder, they use variants of RNN.
For our experiments, we use a pre-trained bidirectional LSTM with max pooling of the final hidden state of both the forward and backward LSTMs.
To train the encoder, inferSent computes concatenation, element-wise multiplication and subtraction of sentence pairs.
These values are then passed through classifier to classify the sentence pairs under 3 categories: entailment, contradiction and neutral.

\textbf{Universal sentence encoder}~\cite{cer2018universal} encodes text into high dimensional vectors using transformers~\cite{vaswani2017attention}.
The model is trained on a variety of data of different lengths, over a variety of tasks.
We use the pre-trained universal-sentence-encoder-large model for our experiments.
It is trained using transformer encoders that use self-attention to compute context aware representation of words in a sentence.
These representations are then turned into fixed length sentence encoding by computing their element-wise sum.
The final representation is formed by dividing this vector with the square root of its length to reduce the effect of sentence length.

\textbf{SentenceBert}~\cite{reimers2019sbert} is another sentence embedding technique that uses a siamese network architecture to generate fixed length representation of sentences.
It is similar to InferSent model architecture, with the main difference being that it uses BERT~\cite{devlin2019bert} instead of LSTM to output the encoding.
For our experiment, we use the stsb-roberta-large pre-trained model~\cite{liu2019roberta}.
Similar to the model we use in BertScore, SentenceBert uses RoBERTa$_{large}$ to generate embeddings for the words in a sentence.
The model then mean-pools these word embeddings to generate a sentence embedding.

\textbf{Attendgru} is an encoder-decoder based neural machine translation model~\cite{luong2015effective}.
Both the encoder and decoder layers of the attendgru model uses a GRU~\cite{cho2014learning} to encode the text.
The model uses the final hidden state of the encoder layer as the initial state of the decoder layer.
For our experiments, we use an attendgru model that was trained on the task of source code summarization~\cite{leclair2019neural}.
During training, the encoder takes source code text and the decoder takes the corresponding comment as input.
We train this model on the java dataset of 2.1m code-comment pairs discussed in section~\ref{dataset}.
We then extract the decoder layer from the trained model and pass the sentences through it to obtain a word encoding.
Finally, we flatten this encoding to obtain a sentence-level encoding of the reference and predicted output.

%% file: corrstudy.tex
\newpage
\section{Quantitative Experiment}
\label{sec:semexp}

This section describes our quantitative experiment in which we study the correlation between the text similarity metrics described in Section~\ref{sec:simmetric} and the human ratings we collected in Section~\ref{sec:humexp}.

\vspace{-1mm}
\subsection{Research Questions}

The research objective of this experiment is to determine which similarity metric most-closely represents human perception of similarity of source code summaries, as a guide for future research in source code summarization.  Our RQs are:

\begin{description}
	\item[RQ4] Which similarity metric most-closely correlates with the human expert ratings for similarity?
	\item[RQ5] What is the correlation of the similarity metrics to each other and to human ratings?
\end{description}

The rationale for RQ4 is that many similarity metrics (and text embeddings for use in similarity metrics) have been proposed, yet the literature does not provide clear guidance on which should be preferred for source code summarization research.  The dominant metric is currently BLEU, and weaknesses of BLEU have been highlighted for several years~\cite{stapleton2020human}, though it is not clear if any existing metric should replace it.

The rationale for RQ5 is that different metrics may measure different aspects of the summaries.  Recall we asked human experts to rate the quality of the summaries in terms of accuracy, completeness, and conciseness in addition to similarity.  The goal of this RQ is to determine the degree of correlation between similarity metrics and these measures, as well as to determine if those metrics tend to correlate with each other.  If some metrics correlate closely with each other, they could be useful alternatives if another is not available.  If some metrics correlate to accuracy, completeness, or conciseness, they may be suitable proxies for those quality criteria.

\vspace{-1mm}
\subsection{Methodology}

Our methodology for answering RQ4 is two-fold.  First, we compute the Spearman Rank Correlation between each similarity metric and the human-ratings for similarity (as well as human ratings for other quality criteria).  The Spearman rank correlation test is appropriate namely because it is nonparametric.  We do not assume a normal distribution due to limited dataset sizes, and therefore do not select a parametric test such as Pearson correlation.  Note that we report the ``rho'' value for Spearman correlation but not the p-value.  As Kay~\cite{kay2009wanted} points out, the p-value of correlation can be misleading because it may give a higher confidence than warranted.

However, we report an alternative p-value to provide a secondary, corroborative view of the data.  The procedure is:

1) We divide the source code summaries into two groups.  One group has summaries for which the aggregate human rating for similarity (or accuracy, completeness, conciseness) is $<=$2.  These correspond to summaries for which the participants averaged ``Strongly Agree'' or ``Agree.''  The second group has summaries where the human rating averaged $>=$3.  These correspond to summaries with an average of ``Disagree'' or ``Strongly Disagree.''

2) We perform a Mann-Whitney U test on the similarity metrics' scores between the groups.  For example, to compare BLEU scores for the $<=$2 (``Agree'') group to the BLEU scores for the $>=$3 (``Disagree'') group.  Then Mann-Whitney test is appropriate because it is unpaired and nonparametric.  We do not assume a normal distribution, and we do not assume a monotonic relationship between items in the groups.  The p-value from this test is a measure of how separated the two groups are.  A very low p-value would indicate a more significant difference in the similarity metric's measurement of the ``Agree'' and ``Disagree'' groups.

Our methodology for answering RQ5 is similar to RQ4, except that we compute the Spearman correlation between every pair of metrics and human ratings.  In addition, we compute Kendall tau correlation between every pair, to provide an alternative view.

\vspace{-2mm}
\subsection{Threats to Validity}

The key threats to validity to this study are similar to our previous experiment: 1) the ratings provided by human experts, and 2) the selection of subroutines and summaries.  This experiment relies on human expert ratings, which are subjective and may change with different people.  Also, we selected the subroutines from a large dataset randomly, hoping to create a representative sample of the dataset. However, there is still a risk that a different selection could lead to different results.  Furthermore, since the dataset is derived from open-source Java projects, the generalizability beyond open-source Java projects is unknown.

%% file: results.tex
\input restbl




\vspace{-1mm}
\section{Quantitative Experiment Results}

In this section, we answer RQ4 and RQ5 in the context of the results of our quantitative experiment.
\\
\vspace{-5mm}
\subsection{RQ4: Similarity Correlation}

We find that cosine similarity of the Sentence Bert Encoder (SentenceBert+c) and Universal Sentence Encoder (USE+c) sentence representation is the closest to similarity perceived by the human evaluators.  As we note in Table~\ref{tab:corr}, the Spearman correlation between similarity and SentenceBert+c is $0.8228$, and USE+c is $0.8226$, which is traditionally interpreted as a ``strong correlation''~\cite{dancey2007statistics} in studies involving subjective ratings by human experts.  In addition, the Kendall tau correlation of SentenceBert+c is $0.645$ and USE+c is $0.634$, also among the highest scores of any metric we studied to the human ratings (Table~\ref{tab:corr2}).  We note that these levels of correlation is markedly higher than BLEU (which reached $0.7191$ Spearman correlation and $0.544$ Kendall tau correlation), implying that SentenceBert+c and USE+c is a better indicator of similarity for source code summaries than BLEU.  In fact, a majority of the metrics we studied, including other n-gram-based metrics, had higher correlation to the human expert ratings of similarity than BLEU.  These findings broadly align with experiments evaluating BLEU by Stapleton~\emph{et al.}~\cite{stapleton2020human}.

A majority of the levels of correlation to human-rated similarity are in the $0.7-0.8$ range.  SentenceBert and USE are slightly higher and BLEU is slightly lower.  The similarities based on attendgru's decoder are far lower, which may seem surprising considering that they were trained using the code summaries themselves.  On the other hand, models such as SentenceBert, USE, and BERTScore were trained on very large natural language datasets, and so are likely to include knowledge of the meaning of words in these datasets, beyond the scope of only code summaries.  Since human readers will also know the meaning of these words, models like SentenceBert and USE are likely able to better represent what humans expect.

\subsection{RQ5: Other Correlations}

We find that both USE+c and SentenceBert+c are also the two strongest overall performers in terms of correlation to human ratings for accuracy and completeness, though these levels of correlation are lower across the board.  SentenceBert+c has one of the highest correlation to accuracy, at $0.5228$ Spearman correlation which is slightly lower than the $0.5242$ achieved by USE+c and $0.5286$ achieved by euclidean distance of SentenceBert (SentenceBert+e). It also has one of the highest correlation to completeness, at $0.5370$, which is again only slightly lower than the $0.5482$ achieved by USE+c.  In general, though, correlation in the $0.4-0.5$ range is only considered in the low end of moderate~\cite{dancey2007statistics}.  These scores are still markedly higher than BLEU, which reaches correlation around $0.450$ for accuracy and $0.478$ for completeness.

\input cmtbl

The lower overall correlation scores to accuracy and completeness probably reflect problems in the reference summaries themselves.  As we noted in our discussion of RQ2, the reference summaries are not necessarily a ``gold standard'' even after substantial filtering.  Since the vast majority of neural code summarization approaches are trained to duplicate these references, they will tend to reflect the problems that exist in these references.  Still, we do note that higher levels of correlation to similarity are associated with higher levels of correlation to accuracy: SentenceBert has the highest correlation to human-rated similarity and near the highest for accuracy, while BLEU has near the lowest for both.  Therefore, even though the reference summaries have problems, it still does seem worthwhile to measure similarity to them.

Correlation to conciseness is relatively low for all metrics, reaching only a maximum of $0.3887$ for ROUGE-W. It reaches only a score of $0.3128$ for BLEU.  The low score for BLEU may be at least partially explained by BLEU's brevity penalty.  Short predictions are penalized by BLEU in an effort to prevent ``gaming'' of BLEU scores by models which just predict very short sentences.  However, short predictions are also more likely to be rated as concise.  Even so, the predictive power of any metric we study to the quality criterion conciseness is low.  The highest level of correlation to conciseness is only considered ``weak''~\cite{dancey2007statistics}, and higher levels of correlation to conciseness do not seem associated with higher levels of correlation to similarity.


The corroborating evidence presented by the p-values in Table~\ref{tab:corr} serve as an important ``sanity check'' to the correlation numbers. Recall that these p-values are for a difference test between groups of ratings where the aggregate human evaluator score is $<=$2 indicating Agree/Strongly Agree (or the aggregate human evaluator score is $>=$3 indicating Disagree/Strongly Disagree). All approaches, including BLEU, received very low p-values for this test, indicating at least a broad pattern of these metrics being associated with similarity. The scores based on attendgru were the highest, which mirrors the relatively low correlation of this metric to accuracy, completeness and conciseness.

Three observations are evident in Table~\ref{tab:corr2}.  First, the Spearman correlations are generally in agreement with the Kendall correlations, regarding which metrics are the most or least correlated with the ratings from human experts.  These correlations are calculated differently and cannot be directly compared, but the relative rank of the metrics to each other is about the same.  Second, correlation to human-rated similarity does not necessarily preclude correlation to BLEU.  For instance, USE+c, InferSent and BERTScore, are all more correlated to BLEU than SentenceBert is, even though SentenceBert is more correlated to human-rated similarity than BLEU.  Finally, correlation of human-rated similarity is only moderate to the quality metrics accuracy and completeness, even though these quality metrics seem to be highly correlated to each other.  
Further study is needed to determine if these perceptions indicate problems in the reference summaries, or merely reflect a tendency to be ``strict'' when rating accuracy.

%% file: restbl.tex
\begin{table*}[!t]
	\centering
	\caption{Correlation and p-values of difference tests of similarity, accuracy, completeness, and conciseness ratings to metrics we study.}
	\label{tab:corr}
\begin{tabular}{lllllll|llll}
	&                  &                          & \multicolumn{4}{c|}{correlation}                                                                              & \multicolumn{4}{c}{p-value of diff. test}                                                                    \\
	&                  &                          & \multicolumn{1}{c}{sim.} & \multicolumn{1}{c}{accu.} & \multicolumn{1}{c}{comp.} & \multicolumn{1}{c|}{cons.} & \multicolumn{1}{c}{sim.} & \multicolumn{1}{c}{accu.} & \multicolumn{1}{c}{comp.} & \multicolumn{1}{c}{cons.} \\ \cline{4-11} 
	n-gram based:           & BLEU             & \multicolumn{1}{l|}{B1}  & 0.7375 & 0.4612 & 0.5124 & 0.3206 
																		  & $<$0.001 & $<$0.001 & $<$0.001 & $<$0.001 \\
	&                  & \multicolumn{1}{l|}{BA}  & 0.7190 & 0.4496 & 0.4783 & 0.3128 
												  & $<$0.001 & $<$0.001 & $<$0.001 & $<$0.001 \\
	& ROUGE            & \multicolumn{1}{l|}{LCS} & 0.8002 & 0.5081 & 0.4730 & 0.3718 
												  & $<$0.001 & $<$0.001 & $<$0.001 & $<$0.001 \\
	&                  & \multicolumn{1}{l|}{W}   & 0.7833 & 0.5106 & 0.4539 & \textbf{0.3887} 
												  & $<$0.001 & $<$0.001 & $<$0.001 & $<$0.001 \\
	& METEOR           & \multicolumn{1}{l|}{}    & 0.7510 & 0.4736 & 0.4715 & 0.3387 
												  & $<$0.001 & $<$0.001 & $<$0.001 & $<$0.001 \\
	& Jaccard          & \multicolumn{1}{l|}{}    & 0.7586 & 0.4774 & 0.4792 & 0.3323 
												  & $<$0.001 & $<$0.001 & $<$0.001 & $<$0.001 \\ \cline{2-11}
	embedding based: & BERTScore        & \multicolumn{1}{l|}{}     & 0.7670 & 0.5037 & 0.4985 & 0.3626 
																	& $<$0.001 & $<$0.001 & $<$0.001 & $<$0.001 \\
	& TF/IDF           & \multicolumn{1}{l|}{c}   & 0.6962 & 0.4312 & 0.4316 & 0.3032 
												  & $<$0.001 & $<$0.001 & $<$0.001 & $<$0.001 \\
	&                  & \multicolumn{1}{l|}{e}   & 0.6923 & 0.4238 & 0.4240 & 0.2945 
												  & $<$0.001 & $<$0.001 & $<$0.001 & $<$0.001 \\
	& InferSent        & \multicolumn{1}{l|}{c}   & 0.7506 & 0.5011 & 0.5084 & 0.3088 
												  & $<$0.001 & $<$0.001 & $<$0.001 & $<$0.001 \\
	&                  & \multicolumn{1}{l|}{e}   & 0.8073 & 0.5053 & 0.5111 & 0.3185 
												  & $<$0.001 & $<$0.001 & $<$0.001 & $<$0.001 \\
	& Univ. Sent. Enc. & \multicolumn{1}{l|}{c}   & 0.8226 & 0.5242 & \textbf{0.5482} & 0.3309 
												  & $<$0.001 & $<$0.001 & $<$0.001 & $<$0.001 \\
	&                  & \multicolumn{1}{l|}{e}   & 0.8201 & 0.5176 & 0.5415 & 0.3223 
												  & $<$0.001 & $<$0.001 & $<$0.001 & $<$0.001 \\
	& SentenceBert     & \multicolumn{1}{l|}{c}   & \textbf{0.8228} & 0.5228 & 0.5370 & 0.3562 
												  & $<$0.001 & $<$0.001 & $<$0.001 & $<$0.001 \\
	&                  & \multicolumn{1}{l|}{e}   & 0.8207 & \textbf{0.5286} & 0.5331 & 0.3589 
												  & $<$0.001 & $<$0.001 & $<$0.001 & $<$0.001 \\
	& attendgru        & \multicolumn{1}{l|}{c}   & 0.5600 & 0.3486 & 0.3716 & 0.2209 
												  & $<$0.001 & $<$0.001 & $<$0.001 & $<$0.001 \\
	&                  & \multicolumn{1}{l|}{e}   & 0.5702 & 0.3483 & 0.3930 & 0.2123 
												  & $<$0.001 & $<$0.001 & $<$0.001 & $<$0.001 
\end{tabular}
\end{table*}

%% file: cmtbl.tex
\newcommand\RotText[1]{
	\rotatebox[origin=c]{90}{\parbox{1.5cm}{%
			#1}}}

\begin{table*}[!t]
	\renewcommand{\arraystretch}{1.5}
	\centering
	\caption{Correlation between different metrics and human ratings in our experiment.  The upper triangle shows Kendall tau correlation, while the lower triangle shows Spearman's rho correlation.  The last four items are the ratings by human experts.}
	\label{tab:corr2}
\begin{tabular}{lp{0.5cm}p{0.5cm}p{0.5cm}p{0.5cm}p{0.5cm}p{0.5cm}p{0.5cm}p{0.5cm}p{0.5cm}p{0.5cm}p{0.5cm}p{0.5cm}|p{0.5cm}p{0.5cm}p{0.5cm}p{0.5cm}}
             & \RotText{B1} & \RotText{BA} & \RotText{R-LCS} & \RotText{R-W} & \RotText{METEOR} & \RotText{Jaccard} & \RotText{BERTScore} & \RotText{TF/IDF} & \RotText{InterSent} & \RotText{U.S.E.} & \RotText{Sent. BERT} & \RotText{attendgru} & \RotText{similarity} & \RotText{accuracy} & \RotText{completeness} & \RotText{conciseness} \\
B1           & 1.000 & 0.790 & 0.754 & 0.721 & 0.729 & 0.836 & 0.633 & 0.774 & 0.671 & 0.654 & 0.547 & 0.437 & 0.560 & 0.327 & 0.362 & 0.229 \\
BA           & 0.922 & 1.000 & 0.707 & 0.692 & 0.731 & 0.747 & 0.613 & 0.687 & 0.624 & 0.616 & 0.535 & 0.393 & 0.544 & 0.321 & 0.339 & 0.221 \\
R-LCS        & 0.900 & 0.869 & 1.000 & 0.927 & 0.800 & 0.807 & 0.636 & 0.760 & 0.649 & 0.677 & 0.590 & 0.408 & 0.621 & 0.365 & 0.333 & 0.268 \\
R-W          & 0.878 & 0.856 & 0.991 & 1.000 & 0.765 & 0.791 & 0.621 & 0.747 & 0.622 & 0.639 & 0.565 & 0.401 & 0.606 & 0.366 & 0.319 & 0.281 \\
METEOR       & 0.877 & 0.887 & 0.932 & 0.914 & 1.000 & 0.740 & 0.593 & 0.664 & 0.607 & 0.644 & 0.552 & 0.375 & 0.577 & 0.336 & 0.335 & 0.240 \\
Jaccard      & 0.944 & 0.892 & 0.931 & 0.927 & 0.878 & 1.000 & 0.633 & 0.831 & 0.682 & 0.655 & 0.570 & 0.376 & 0.582 & 0.341 & 0.346 & 0.237 \\
BERTScore    & 0.800 & 0.785 & 0.811 & 0.795 & 0.771 & 0.800 & 1.000 & 0.596 & 0.594 & 0.619 & 0.573 & 0.420 & 0.593 & 0.360 & 0.352 & 0.254 \\
TF/IDF       & 0.912 & 0.846 & 0.905 & 0.900 & 0.819 & 0.945 & 0.764 & 1.000 & 0.663 & 0.625 & 0.515 & 0.340 & 0.522 & 0.308 & 0.309 & 0.213 \\
InterSent    & 0.836 & 0.799 & 0.818 & 0.795 & 0.775 & 0.846 & 0.770 & 0.835 & 1.000 & 0.658 & 0.595 & 0.279 & 0.561 & 0.349 & 0.360 & 0.215 \\
U.S.E.       & 0.829 & 0.790 & 0.849 & 0.817 & 0.820 & 0.828 & 0.800 & 0.798 & 0.834 & 1.000 & 0.632 & 0.404 & 0.634 & 0.368 & 0.385 & 0.229 \\
Sent. BERT   & 0.735 & 0.721 & 0.768 & 0.740 & 0.735 & 0.751 & 0.767 & 0.695 & 0.778 & 0.815 & 1.000 & 0.368 & 0.645 & 0.373 & 0.381 & 0.248 \\
attendgru    & 0.586 & 0.531 & 0.558 & 0.556 & 0.522 & 0.512 & 0.588 & 0.476 & 0.394 & 0.547 & 0.517 & 1.000 & 0.402 & 0.246 & 0.261 & 0.154 \\ \cline{1-17}
similarity   & 0.738 & 0.719 & 0.800 & 0.783 & 0.751 & 0.759 & 0.767 & 0.696 & 0.751 & 0.823 & 0.823 & 0.560 & 1.000 & 0.718 & 0.690 & 0.510 \\
accuracy     & 0.461 & 0.450 & 0.508 & 0.511 & 0.474 & 0.477 & 0.504 & 0.431 & 0.501 & 0.524 & 0.523 & 0.349 & 0.549 & 1.000 & 0.713 & 0.575 \\
completeness & 0.512 & 0.478 & 0.473 & 0.454 & 0.472 & 0.479 & 0.498 & 0.432 & 0.508 & 0.548 & 0.537 & 0.372 & 0.528 & 0.873 & 1.000 & 0.462 \\
conciseness  & 0.321 & 0.313 & 0.372 & 0.389 & 0.339 & 0.332 & 0.363 & 0.303 & 0.309 & 0.331 & 0.356 & 0.221 & 0.366 & 0.757 & 0.628 & 1.000   
\end{tabular}
\end{table*}

%% file: discussion.tex
\vspace{-0.2cm}
\section{Discussion \& Conclusions}
\label{sec:reproducibility}

This paper moves the state-of-the-art forward by showing how the predominant procedure for evaluating current source code summarization techniques (BLEU score) may be superseded by other metrics.  
Most source code summarization techniques are trained to mimic the summaries that exist in large code repositories.  
They are then evaluated using automated metrics that compare the predicted summaries to the reference summaries. There are many advantages to using an automated metric, such as the ability to compare many thousands of examples. On the other hand, the metric used to compute similarity is a weakpoint.  The current, predominant metric is BLEU.  In this paper, we find that other metrics better represent perceptions of similarity by human evaluators at the sentence-level.

Our recommendations to future researchers are:
\begin{enumerate}
	\item Compute the Sentence Bert Encoder + cosine similarity metric when comparing predicted source code summaries to reference summaries at the sentence-level.  This metric accounts for semantic similarity of summaries, and was the highest correlated to human-rated similarity in this paper. 
	\item Continue to report corpus-level BLEU scores. 
	BLEU retains value in consistency with related work, and the results of this paper (at the sentence-level) should not be considered generalizable to corpus-level results.  However, researchers are advised to consider warnings regarding corpus-level BLEU~\cite{roy2021reassessing}.
	\item If a human study is feasible, consider using the quality criteria accuracy, completeness, and conciseness, to be consistent with related work.  Also, consider reporting correlation of these ratings to SentenceBert+c and BLEU.
\end{enumerate}

This paper adds to a growing consensus that BLEU alone is not a sufficient means to evaluate source code summarization techniques.  While automated procedures are desirable due to low cost and relative ease of reproducibility when compared to human studies, BLEU as a sentence-level metric is not as well correlated to human ratings of similarity as other metrics.  
Related papers find that BLEU is not associated with better program comprehension~\cite{stapleton2020human}, and suggest focusing on key parts of summaries such as the action words~\cite{haque2021action}. 
Our contribution is to show that while BLEU retains value as a means of consistency with previous work, newer metrics are poised to replace it in automated evaluation procedures for code summarization.

We provide the following online appendix for reproducibility.

{\scriptsize \texttt{https://github.com/similarityMetrics/similarityMetrics}}
